\documentclass[amsmath,amssymb]{revtex4}

\usepackage{graphicx}
\usepackage{epsfig}
\usepackage{dcolumn}
\usepackage{bm}

\newcommand {\bean}  {\begin{eqnarray*}}
\newcommand {\eean}  {\end{eqnarray*}}

\def\Rset {{\rm I \kern-.2em R}} 

\newcommand {\ds}   {\displaystyle}

\newcommand {\bce}  {\begin{center}}
\newcommand {\ece}  {\end{center}}
\newcommand {\be}   {\begin{equation}}
\newcommand {\ba}   {\begin{array}}
\newcommand {\bea}  {\begin{eqnarray}}
\newcommand {\bfi}  {\begin{figure}}
\newcommand {\ee}   {\end{equation}} 
\newcommand {\ea}   {\end{array}}
\newcommand {\eea}  {\end{eqnarray}}
\newcommand {\efi}  {\end{figure}}

\newcommand {\noi}  {\noindent}

\newcommand {\UNIV}   {Universit\`a }

\def\Rset {{\rm I \kern-.2em R}} 
\def\mathbbH {{\rm I \kern-.2em H}} 
\def\mathbbC {{\rm I \kern-.6em C}} 

\newcommand\msec{\nobreak\mbox{$\;$m\,s$^{-1}$}}
\newcommand\micron{\nobreak\mbox{$\,$ $\mu\,$m}}
\newcommand\m{\nobreak\mbox{$\;$m}}
\newcommand\mm{\nobreak\mbox{$\;$mm}}

\newcommand\kHz{\nobreak\mbox{$\;$kHz}}

\newcommand{\la}{\left\langle}
\newcommand{\ra}{\right\rangle}

\newcommand{\x}{\mbox{\boldmath$x$}}
\renewcommand{\r}{\mbox{\boldmath$r$}}

\begin{document}

\preprint{APS/123-QED}

\title{Anisotropic fluctuations in turbulent sheared flows}

\author{Boris Jacob$^1$}

\author{Luca Biferale$^2$}

\author{Gaetano Iuso$^3$}

\author{Carlo Massimo Casciola$^1$}

\affiliation{$1$ Dip. Mecc. Aeron., \UNIV di Roma ``La Sapienza'',
             via Eudossiana 18, 00184 Roma, Italy.}

\affiliation{$2$ Dip. di Fisica, and INFM \UNIV ``Tor Vergata'',
        via della Ricerca Scientifica 1, 00133 Roma, Italy.}

\affiliation{$3$ DIAS, Politecnico di Torino,
             corso Duca degli Abruzzi 24, 10129 Torino, Italy.}

\date{\today}

%
%
%
%
%
%

\begin{abstract}
An experimental analysis of small-scales anisotropic turbulent
fluctuations has been performed in two different flows. We analyzed
anisotropic properties of an homogeneous shear flows and of a turbulent
boundary layer by means of two cross-wire probes to obtain multi-point
multi-component measurements. Data are analyzed at changing inter-probe 
separation  without the use of Taylor hypothesis. The results are consistent 
with the ``exponent-only'' scenario for universality, i.e. 
all experimental data can be fit by fixing the same set of anisotropic scaling
exponents at changing only prefactors, for different shear
intensities and boundary conditions.
\end{abstract}


\maketitle

%
%
%
%
%

\section{Introduction} 
Statistical theory of turbulence is often focused on homogeneous and
isotropic flows \cite{fri95}.  Experimentally, however, we know that
isotropy holds approximately, with different degrees of
justification, depending on the geometry of the boundaries
and on the driving mechanism.  Therefore, a realistic description of
turbulence cannot ignore anisotropic and non-homogeneous effects,
especially in regions close to the boundaries, and/or at scales close to  the integral
scale, $L_0$, where the injection mechanism can strongly affect
velocity fluctuations. The interest of quantifying anisotropic and
non-homogeneous effects  is also linked to the
important issue of ``recovery of isotropy'', i.e. the problem of
``small-scales universality''. 
Surprisingly enough, recent experimental and numerical
works \cite{gar98,she00,she02a,pum96,bif01a,bif01} have detected
the survival of anisotropic turbulent fluctuations till to 
the Kolmogorov scale, $\eta$.
These findings have stimulated a lot of further experimental,
numerical and theoretical work focused to develop proper analytical
tools \cite{ara99b}
and to extend the available experimental/numerical data sets 
\cite{gar98,she00,she02a,pum96,ara99,bif01a,bif01,ara98,kur00,kur00a}. Many progresses have been done. For example,
the so-called puzzle of ``persistence of anisotropies'' in
small-scales --high Reynolds numbers-- sheared flows, has been 
recently understood as the effect of the presence of strong anomalous
anisotropic fluctuations \cite{bif01,bif02}.  The attention is mainly focused 
 on correlation functions  based on gradients  (to probe 
Reynolds number dependencies) or to the projections on
isotropic/anisotropic sectors of multi-points velocity correlation
functions, $S^{\{\alpha\}}(\r_1,\dots,\r_n)
\equiv \la v_{\alpha_1}(\r_1)\dots v_{\alpha_n}(\r_n)\ra $, where we use 
 $\{\alpha\}$ as a shorthand notation for the ensemble of indexes
 $\alpha_1,\dots,\alpha_n$ .  When all spatial separation, $\r_1 \dots \r_n$,
 are  in the inertial range, $\eta \ll |\r_i -\r_j|
\sim r \ll L_0$, one expects the existence of power laws behavior under a 
uniform space dilation: $S^{\{\alpha\}}(\lambda
\r_1,\dots,\lambda\r_n)
=\lambda^{\xi(n)}S^{\{\alpha\}}(\r_1,\dots,\r_n)$. Most of the recent
works in anisotropic turbulence  concentrated on 
determining the values of the exponent, $\xi(n)$, as a function of the
order of the correlation function, $n$, and of its anisotropic
properties. Indeed, an important step forward has been done by
realizing that different projections of the multi-point correlation
functions on different irreducible representation of the group of
rotation, SO(3), possess different scaling properties. The idea is to
decompose any  correlation function in a complete basis of eigenfunctions
with defined properties under rotations. Each eigenfunction 
identifies a specific anisotropic sector with total angular momentum, $j$,  
and  its projection, $m$, on a given axis. 
It is believed that the scaling properties of the projections on different 
sectors  possess different scaling exponents, $\xi_j(n)$.

Exponents for the fully isotropic sectors are labeled by $j=0$ while
more and more anisotropic fluctuations are measured by higher and higher  
values of  $j$  \cite{ara99b,ara98}.
 
The higher-than-expected presence of small-scales anisotropic fluctuations 
raises questions about their  universal or non-universal origins. 
In other words, one is interested to control if {\it all} flows posses the 
same, or similar, anisotropic small-scales fluctuations independently on 
their large-scale behavior. 
Of course, full universality cannot be expected, one is tempted to believed  
to a ``exponent-only'' scenario, i.e. only the scaling exponents, $\xi_j(n)$, 
pertaining to each different anisotropic sector, are universal, while the 
overall correlation functions intensities are not. 
This hypothesis is inspired by both theoretical reasons \cite{ara99b,dec} 
and similarities with 
what observed for isotropic fluctuations \cite{arneodo}. 
	
Up to now, there are a few experimental and numerical data sets where  
universality of anisotropic fluctuations  has been probed. 
In particular, as of today, we have some detailed experimental investigation 
of anisotropic small-scale fluctuations in homogeneous shear flows
\cite{she00,she02a}, in atmospheric boundary layer \cite{ara98,kur00,kur00a} 
and in windtunnel data \cite{wandervater}.  
From the numerical side, only a few  Direct Numerical Simulations
 (DNS) in highly 
anisotropic flows have been performed with the aim to explicitly  test 
the small-scales properties of anisotropic fluctuations \cite{ara99,Mazzitelli,bif01a,bif03a,bif02}. 
The situation is still moot.  On the  experimental side, because of the 
difficulty to have multi-point multi-components velocity measurements one can
access only the $j=2$ sector.
On the other hand, DNS can properly disentangle fluctuations of all sectors, 
but due to limitations in the Reynolds numbers, only results in the $j=4$ 
and $j=6$ sectors have been obtained with some accuracy.  
The $j=2$ sector in the numerical works 
\cite{bif01a,bif03a} was not measurable due to strong finite Reynolds 
effects.  
Results from different experiments, with different geometries and different 
large scale structures, are in fairly good agreement concerning the 
$j=2$ sector up to moment $n=6$.  
Putting together all results of numerics and experiments one 
recovers a scenario for anisotropic fluctuations consistent (not in 
contradiction)  with the ``exponent-only'' picture of universality.
Still, more tests in both experiments and numerics are needed.

The aim of this paper is to present a new systematic assessment of
anisotropic fluctuations in sheared flows at changing
both the experimental set-up and the shear intensity.  In particular,
we have measured small-scales turbulent properties in a
homogeneous shear flow (HS) and in a turbulent boundary layer (TBL).
One of the novelties here presented consists in the using of
measurements from two cross-wire probes at changing their 
separation, i.e. we do not need to use Taylor hypothesis of frozen turbulence
in order to extract information at different scales. 
This kind of multi-points multi-components measurements are necessary in order 
to disentangle contributions from isotropic and anisotropic
fluctuations and among  different kind of anisotropic fluctuations.\\ 
Our results  support  the ``exponent-only'' scenario. We found
good qualitative, and quantitative, agreement of the anisotropic
scaling exponents in both HS and TBL flows. Moreover our results
are in agreement with the previously measured values in different experiments 
with different Reynolds numbers and different shear intensities. \\ 
The paper is organized as follows. 
In sec.~\ref{apparatus} we present the details of the two experimental 
apparatus including some typical large-scale measurements to validate the 
laboratory set-up. 
In sec.~\ref{results} we present the scaling properties for both HS and TBL 
flows. Conclusions follow in sec.~\ref{concluding}. 

%
%
%
%

\section{Experimental set-up} 
\label{apparatus} 

The data we are going to   discuss concern two different experiments, both 
conducted in the $1.30 \m \times 0.90 \m$ test section of a 7 m long open return 
wind tunnel.
The first data-set has been obtained in a nominally homogeneous shear flow
(HS), 
characterized by a constant velocity gradient.  
The second one refers to  measurements performed in the logarithmic 
region of a zero-pressure gradient turbulent boundary layer (TBL).


\subsection{Homogeneous shear flow}

\begin{figure}[h]
{\centerline {
\epsfig{figure=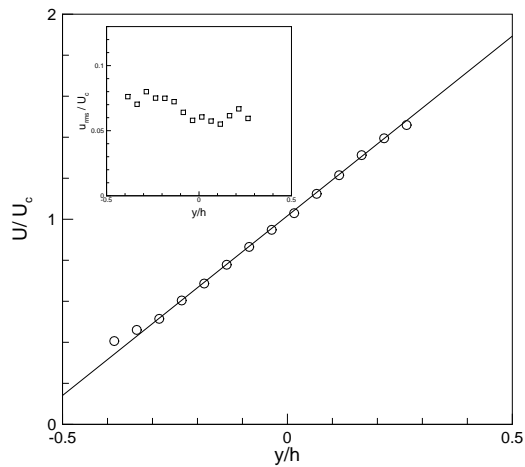,width=8.0cm}}}
   \caption{Homogeneous shear flow:
            mean velocity profile as a function of the non-dimensional
            coordinate in the mean gradient direction, $y/h$.
            In the inset, the normalized streamwise turbulent intensity
            $u_{\rm rms}$.
            Velocities are normalized in terms of the centerline velocity,
            $U_c = 10.2 \, \msec$, while the test section height is
            $h=0.9 \,\m$.
            \label{HSF}
                        }
\end{figure}

\begin{table}
\begin{ruledtabular}
\begin{tabular}{cccccccc}
$u_{\rm rms}$  & 
$v_{\rm rms}$  & 
$\rho_{uv}$    & 
$\epsilon$     & 
$\ell_{\rm T}$            & 
$\eta$         & 
$Re_{\lambda}$ & 
$S^*$          \\
(\msec)                         & 
(\msec)                         & 
 $ $                            & 
($\;{\rm m}^{2}\,{\rm s}^{-3}$) & 
(\mm)                           &
(\mm)                           &
 $ $                            &
 $ $                            \\
\hline
0.43 & 
0.31 & 
-0.29 & 
0.6 & 
53 & 
0.28 & 
170 & 
4.9 \\
\end{tabular}
\end{ruledtabular}
\caption{\label{tab:HS-table} 
Basic parameters for the homogeneous shear flow. Symbols are defined as 
follows: The rms fluctuation intensity in the streamwise direction is  
$u_{\rm rms} = {\langle ( u - U )^2 \rangle}^{1/2}$, and analogous definitions
apply for $v_{\rm rms}$ and $w_{\rm rms}$.
$\rho_{uv} = \langle u v \rangle/(u_{\rm rms} \, v_{\rm rms})$ is the 
correlation coefficient, $\epsilon$ is the energy dissipation rate
evaluated in terms of the one-dimensional spectrum $E_{11}(\kappa_1) $: 
$15 \, \nu  \int \kappa_1^2 E_{11}(\kappa_1) \, d\kappa_1 \, $.
 $\ell_{\rm T}$ is the transverse 
integral length scale, 
$\int \langle \, u(x,y,z) \, u(x,y,z+r_z) \, \rangle \, dr_z/u_{\rm rms}^2$
and $\eta = (\nu^3/\epsilon)^{1/4}$ is the Kolmogorov scale.
The Taylor-Reynolds number is $Re_\lambda = \lambda u_{\rm rms}/\nu$,
where the Taylor scale follows from $\sqrt{15 \nu u_{\rm rms}^2/\epsilon}$
and the shear parameter is $S^* = \sqrt{S u_{\rm rms}^2/\epsilon}$.
}
\end{table}


The set-up  of the homogeneous shear flow is based on the original
idea proposed in  \cite{Tavoularis}.  
The mean shear is generated with  a device consisting of a series of 
$15$ adjacent small channels, equipped with screens of different solidity 
to produce suitable pressure drops. 
The channels are followed by flow straighteners in such a way that 
sufficiently downstream the mean velocity profile 
$U(y)$ is linear in the core region. 
The data shown in figure~\ref{HSF} correspond to a measurement
approximately $4.8 \m$ downstream of the apparatus,  where the flow is 
already well developed.
Concerning fluctuations, the deviations from the ideal constant profile
of the streamwise turbulence intensity 
$u_{\rm rms}(y) = { \, \langle \,  ( u - U )^2\, \rangle \,^{\frac{1}{2}}}$ 
are comparable with what observed in similar set-up \cite{gar98}. 
In particular, in the central part of the test section where the data 
discussed below have been acquired, they are of the order of $7 \, \%$.
The dimensionless shear rate $S^*   \simeq 5$ (table~\ref{tab:HS-table}), 
is a factor two smaller than what achieved in the logarithmic  part of the 
turbulent boundary layer (table~\ref{tab:BL-table}).

\subsection{Boundary layer}

\begin{figure}[h]
{\centerline {
\epsfig{figure=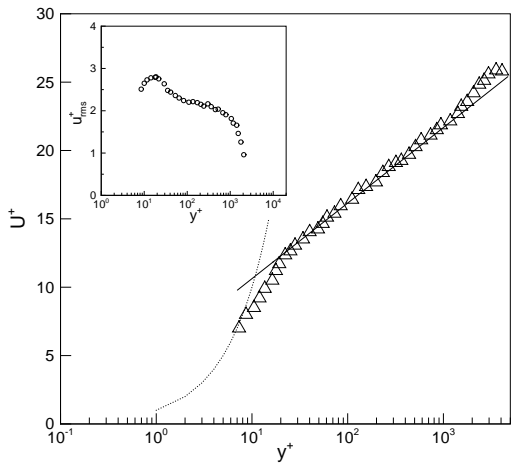,width=8.0cm}}}
 \caption{Turbulent boundary layer: Mean velocity profile $U^+$
          vs wall normal distance $y^+$ in wall units.
          The dotted line corresponds to the linear behavior $U^+=y^+$,
          while the solid line indicates the log-law: $U^+=(1/\kappa) \,
          {\mbox{ln}} \, y^+ +B$, with $\kappa=0.41$ and $B=5.1$.  The
          inset shows the streamwise turbulent fluctuation $u^+_{\rm rms}$.
          Wall units are defined in terms of the friction velocity
          and the kinematic viscosity $\nu$, thus $U^+ = U/u_\tau$,
          $u^+_{\rm rms} = u_{\rm rms}/u_\tau$ and
          $y^+ \, = \, u_\tau \, y/ \nu$.
          \label{profi_tbl}
          }
\end{figure}

The boundary layer develops on the smooth surface of 
the lower  wall of the tunnel, where a nominal zero 
pressure-gradient is achieved by adjusting the upper wall. 
The measurements have been performed 
on the centerline of the test section, $6.0 \m$ downstream the tripping 
device at the end of the contraction. 
With an external velocity $U_\infty$ of  $11.5 \, \rm{m/s}$, the thickness 
of the boundary layer at this location is approximately 
$\delta \simeq 40 \, \rm{mm}$, while the  Reynolds number based on 
the momentum thickness $Re_{\theta}$ is approximately   $6500$, 
well within the range
pertaining to a fully developed turbulent boundary layer. 
The friction velocity, $u_\tau = \sqrt{\tau_w/\rho}$ with $\tau_w$ the average
shear stress at the wall and $\rho$ the constant density,
estimated from the mean velocity profile with a Clauser 
chart, is found to be $u_\tau= 0.43 \msec$,  in good agreement 
($\pm 8 \%$) with a direct measurement by means of a  
Preston tube.
The streamwise mean velocity $U(y)$ and the fluctuation intensity 
$u_{\rm rms}(y)$ profiles are displayed in Figure~\ref{profi_tbl}. 
Both curves show that the flow complies to the requirements of a fully 
developed turbulent boundary layer. 

\begin{table}
\begin{ruledtabular}
\begin{tabular}{ccccccccc}
$y^+$              & 
$u_{\rm rms}$  & 
$v_{\rm rms}$  & 
$\rho_{uv}$    & 
$\epsilon$     & 
$\ell_{\rm T}$            & 
$\eta$         & 
$Re_{\lambda}$ & 
$S^*$          \\
                                &
(\msec)                         & 
(\msec)                         & 
 $ $                            & 
($\;{\rm m}^{2}\,{\rm s}^{-3}$) & 
(\mm)                           &
(\mm)                           &
 $ $                            &
 $ $                            \\
\hline
350 & 
0.94 & 
0.45 & 
-0.37 & 
7.1& 
12.2 & 
0.15 & 
330 & 
12.1 \\
240 & 
1.01 & 
0.47 & 
-0.39 & 
11. & 
8.6. & 
0.13 & 
300 & 
12.8 \\
140 & 
1.06 & 
0.48 & 
-0.38 & 
13.7 & 
7.1  & 
0.125 & 
250 & 
14.2 \\
90 & 
1.07 & 
0.46 & 
-0.36 & 
23. & 
6.6 & 
0.11 & 
230 & 
15.9 \\
\end{tabular}
\end{ruledtabular}
\caption{\label{tab:BL-table} Basic parameters for the boundary layer.
For definitions see caption of table~\ref{tab:HS-table}.}
\end{table} 

%
\subsection{Data acquisition}
%

The instrumentation consists essentially of a couple of 
sub-miniature X-probes, mounted on streamlined supports in order to minimize 
interference effects, and separated in the transverse direction, see 
figure~\ref{2sondeX}.
The wires are  $2.5 \, \micron$ in diameter, $0.5 \mm$ in effective length 
and in separation, oriented at $\pm 45 ^\circ$ with respect to the streamwise 
direction.  They are operated at an overheat ratio 1.9. 
Single component sub-miniature probes (diameter  $2.5 \, \micron$, 
length to diameter ratio equal to 200 to minimize conduction losses) have  
also been used to measure the velocity profiles. 
The  signals from the two X-wires  are simultaneously digitized at 21 $\kHz$ 
with a 16-bit data-acquisition board, after being low-pass filtered at the 
Nyquist frequency and suitably amplified to achieve a good signal-to-noise 
ratio. 
The frequency response of the hot-wires, measured in the free stream at a 
reference velocity $U=6 \, \msec$, is larger than 15 kHz. 
In this way, the  resolution  needed to analyze the small 
scales behavior down to the Kolmogorov length-scale $\eta$ - typically of 
the order of $0.1 \div 0.2 \mm\,$ - is guaranteed.  
The X-wires are calibrated in situ against a Pitot tube. 
The mapping between the output voltages and the components  of the velocity 
(u,v) is obtained by varying both the reference velocities and the orientation 
of the probe with respect to the flow  (see, e.g., \cite{Meneveau}). 
The calibration was repeated at the end of each set of measurements, to check 
that no voltage drift had occurred.   

As for the length of the signals, they  consist  typically of  
$4 \div 8 \times 10^6$ samples, corresponding roughly to $2 \times 10^4$  eddy 
turnover times for both flows.  Convergence of the statistics has been checked
 up to the $6^{\rm th}$ moment.  

%
%
%
%

\section{Results}
\label{results} 
Scaling properties of anisotropic fluctuations are traditionally addressed 
through objects that are identically zero in homogeneous isotropic conditions.
Typically, the study has been confined to the co-spectra
\cite{lumley}.
Recently a new extended set of observables has been proposed in the context 
of the $\rm SO(3)$ decomposition. 
The idea is  to exploit the expansion of any  generic  statistical 
observable in terms of a suitable eigen-basis with a well characterized 
behavior under rotations \cite{she02a,kur00}.
Let us focus, for instance, on the  generic  element in the space of second 
order correlation tensors
\be 
\label{Sab}
S_{\alpha_1 \alpha_2} ({\bf r}) \, = \,
\langle \,  \delta v_{\alpha_1} ({\bf r}) \delta v_{\alpha_2}({\bf r}) \, 
\rangle  \, ,
\ee
\noi where $\delta v_\alpha({\bf r})$ denotes the $\alpha^{\rm th}$ component 
of the velocity increment at two points separated by the vector ${\bf r}$,
$ \delta v_\alpha({\bf r}) \equiv  v_\alpha( \x + \r) -v_\alpha(\x)$.
The appropriate SO(3) decomposition of (\ref{Sab}) reads \cite{ara99b}:
\begin{eqnarray}
\label{SO3}
S_{\alpha_1 \alpha_2}({\bf{r}}) \, = \, 
\sum_{j=0}^{\infty} \,\, \,  
\sum_{m=-j}^{+j} \,\,\,  
\sum_{q=1}^{p(j)}\,\,\, 
S_{jmq}^{(2)}({r})
\, B^{jmq}_{\alpha_1 \alpha_2}(\hat{\bf{r}}) \ .
\nonumber
\end{eqnarray}

\noi Here the index $j$ denotes a sector, to be understood as
a subspace invariant with respect to rotations, 
$B^{jqm}_{\alpha_1 \alpha_2}$ denote the appropriate basis function \cite{ara99b}
which 
depend on the unit vector $\hat{\bf{r}}$ and $p(j)$ counts the number of
irreducible representations.
In particular, $j=0$ labels the isotropic sector, while sectors 
of increasing anisotropy  correspond to higher and higher $j$'s. 
Information on the  dynamics of the system is now captured by the coefficients 
$S_{jmq}^{(2)}$ which depend only on distance $r$.

The invariance under rotations of the inertial terms of the Navier-Stokes
equations suggests that small-scales statistics depend 
only on the sector under consideration.
For Reynolds large enough, scaling laws of the projection are therefore 
expected in the form
$$
S_{jmq}^{(2)}(r) \sim  r^{\xi_{j}(2)},$$ 
where the scaling  exponent explicitly depend on the sector, $j$, while
the argument $2$ reminds that we are presently dealing with a 
second order tensor.
The machinery can be easily extended to structure functions of any order $n$, 
\be 
\label{Sn}
S_{{\alpha_1} \ldots {\alpha_n}} ({\bf r}) \, = \,
\langle \,  \delta v_{\alpha_1} ({\bf r}) \ldots 
\delta v_{\alpha_n} ({\bf r}) \, \rangle  \, ,
\ee

\noi whose projection on the proper SO(3) basis will possess a scaling 
behavior: 
$$S^{\,(n)}_{jmq}(r) \sim  r^{\xi_{j}(n)}.$$
In this context, the recovery of isotropy at smaller and smaller 
scales correspond to  the existence of a hierarchy of exponents
$\xi_{j=0}(n) < \xi_{j}(n)$ \cite{ara99b,bif01}.

\begin{figure}[h]
{\centerline {
\epsfig{figure=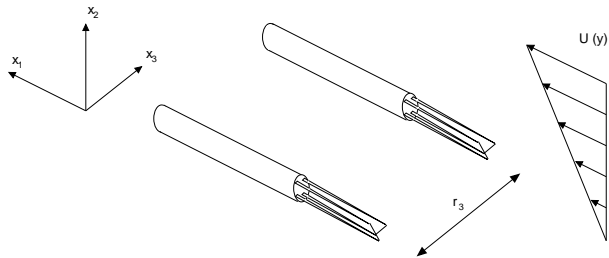,width=8.0cm}}}
   \caption{Schematic of the measurement configuration: each of the two
            X-wires, separated in the transverse direction, $z=x_3$,
            detects two velocity components $u$ and $v$ (in the
            streamwise, $x=x_1$, and in the shear direction $y=x_2$,
            respectively).
            The separation  between the two probes was measured by means of a
            CCD camera with a high magnification lens
            and could be changed by means of a precision transverse gear
            from $0.8 \mm$ to $70 \mm$.
            \label{2sondeX}
}
\end{figure}

Scaling laws for the anisotropic sectors has been recently addressed by 
using different DNS databases \cite{Mazzitelli,bif01,bif01a}.
From the experimental point of view, the evaluation of the proper $\rm SO(3)$
components of a given correlation tensor is hampered by the limited information
on its spatial dependence. 
In the latter  case, the simplest approach is to make a  selection of  
tensorial components such as to cancel out the isotropic contribution in the 
expansion (\ref{SO3}). 
For example, the component  $S_{12}(\r)$ in the direction $\r=(0,0,r_3)$  
vanishes in a  purely isotropic ensemble. Still, in principle all anisotropic 
sectors may influence its behavior. One may follow two possibilities.
Either one may  extract the whole anisotropic spectrum by making a 
multi-parameter fit in all sectors \cite{ara98} or may assume, as done in the 
present paper,  that at scales small enough the correlation function is 
dominated by the leading anisotropic contribution \cite{she02a,kur00}.
Considering that in the geometrical set-up of our interest the $j=1$ sector is
absent by symmetry, one assumes that in the small-scales limit (at high Reynolds numbers) the leading behavior of (\ref{SO3}) is given by the $j=2$ sector:
\begin{eqnarray}
\label{SO3_1}
S_{\alpha_1 \alpha_2}({\bf{r}}) \, \sim \, 
\sum_{m=-2}^{+2} \,\,\,  
\sum_{q=1}^{p(2)}\,\,\, 
S_{2mq}^{(2)}({r})
\, B^{2mq}_{\alpha_1 \alpha_2}(\hat{\bf{r}}) \ .
\nonumber
\end{eqnarray}
Clearly, using  this procedure, systematic, out-of-control, errors are 
introduced by neglecting the higher $j$ sectors. 
Similar consideration can be extended to tensorial correlation functions 
of any order. 
\begin{table}
\begin{ruledtabular}
\begin{tabular}{ccccc}
 n & separation & Observable & \\
\hline
$2$ & $r_1$ & $S_{12}(r_1)$   \\
$4$ & $r_1$ & $S_{1112}(r_1)$  & $S_{1222}(r_1)$ & \\
$6$ & $r_1$ & $S_{111112}(r_1)$ &  $S_{111222}(r_1)$ & $S_{122222}(r_1)$\\
$2$ & $r_3$ & $S_{12}(r_3)$   \\
$4$ & $r_3$ & $S_{1112}(r_3)$  & $S_{1222}(r_3)$ & \\
$6$ & $r_3$ & $S_{111112}(r_3)$ &  $S_{111222}(r_3)$ & $S_{122222}(r_3)$\\
\end{tabular}
\end{ruledtabular}
\caption{\label{tab:Lista_S} List of observables with null contribution
from the $j=0$ and $1$ sectors.
}
\end{table}
For example, table~\ref{tab:Lista_S} lists several observables which, 
according to the previous  discussion  and the symmetries 
of the experimental set-up sketched in figure~\ref{2sondeX}, do not
present contributions  from both sectors $j=0$ and $j=1$.
In the table, the suffixes $1$, $2$ and $3$ correspond to direction
$x_1 = x$,  $x_2 = y$ and $x_3 = z$, respectively. 
The objects reported in the first three lines depend on the streamwise 
separation $r_1$, they can be evaluated by using a single X-wire probe
 and Taylor hypothesis. Those on the last three lines depend on the 
transverse separation $r_3$ and can be measured only by using at least two 
X-wire probes. 
Hereafter we mainly present results based on the two-probes approach.\\

Considering the schematic of figure~\ref{2sondeX}, the two points measurements 
consist in the acquisition of $u$ and $v$ at two points separated in 
direction $z$ (see caption).

On the other hand, for example,  the single point measurement 
of $S_{1112}(r_1)$  with $u_1 = u$ and  $u_2 = v$, as a function of time yields
\bea
\label{Taylor}
S_{1112}(r_1) \, =  
\qquad \qquad \qquad \qquad \qquad \qquad \qquad \qquad 
\nonumber
\\
 \langle \, 
\left[  \, u(t \, + \, r_1/U) \, - \, u(t) \, \right]^3 \, 
\left[  \, v(t \, + \, r_1/U) \, - \, v(t) \, \right] \,  \rangle  \ .
\eea
\noi This approach has been used e.g. in \cite{she02a} in the context of the 
homogeneous shear flow and in \cite{kur00}  at a single location in the 
atmospheric boundary layer to address the scaling properties of the
$j=2$ sector.
In \cite{kur00a}, two single component wires, at fixed separation in the 
transverse direction $z$, are used in connection with Taylor hypothesis to 
extract the scaling exponent of the $j=2$ sector while a similar procedure with
two X-wires separated in the shear direction $y$ permits to investigate the 
scaling behavior of the $j=1$ sector.

Purpose of the present work is to by-pass the use of Taylor hypothesis
   by using
the configuration described in the schematic of figure~\ref{2sondeX}.
This allows to compute the anisotropic observables depending on $r_3$ 
(table~\ref{tab:Lista_S}) by continuously changing the transverse separation 
between the two probes. 

\subsection{The homogeneous shear flow}
\label{subsec:HS} 

The global parameters characterizing the homogeneous shear flow are
summarized in table~\ref{tab:HS-table}. 
In order to allow for the direct comparison of the data for the homogeneous 
shear flow with those for the boundary layer that are presented in 
subsection~\ref{subsec:BL}, a common normalization procedure is used. 
For the homogeneous shear the relevant characteristic velocity is defined as

\bea
\label{u_tau_S}
u_\tau \, = \, \sqrt{ \tau /\rho}
\eea

\noi where the total shear stress is given by 
$ \tau = \nu S \, - \, \langle u v \rangle$ where $S = dU/dy$ is the 
mean shear.

\begin{figure}[h]
{\centerline {
\epsfig{figure=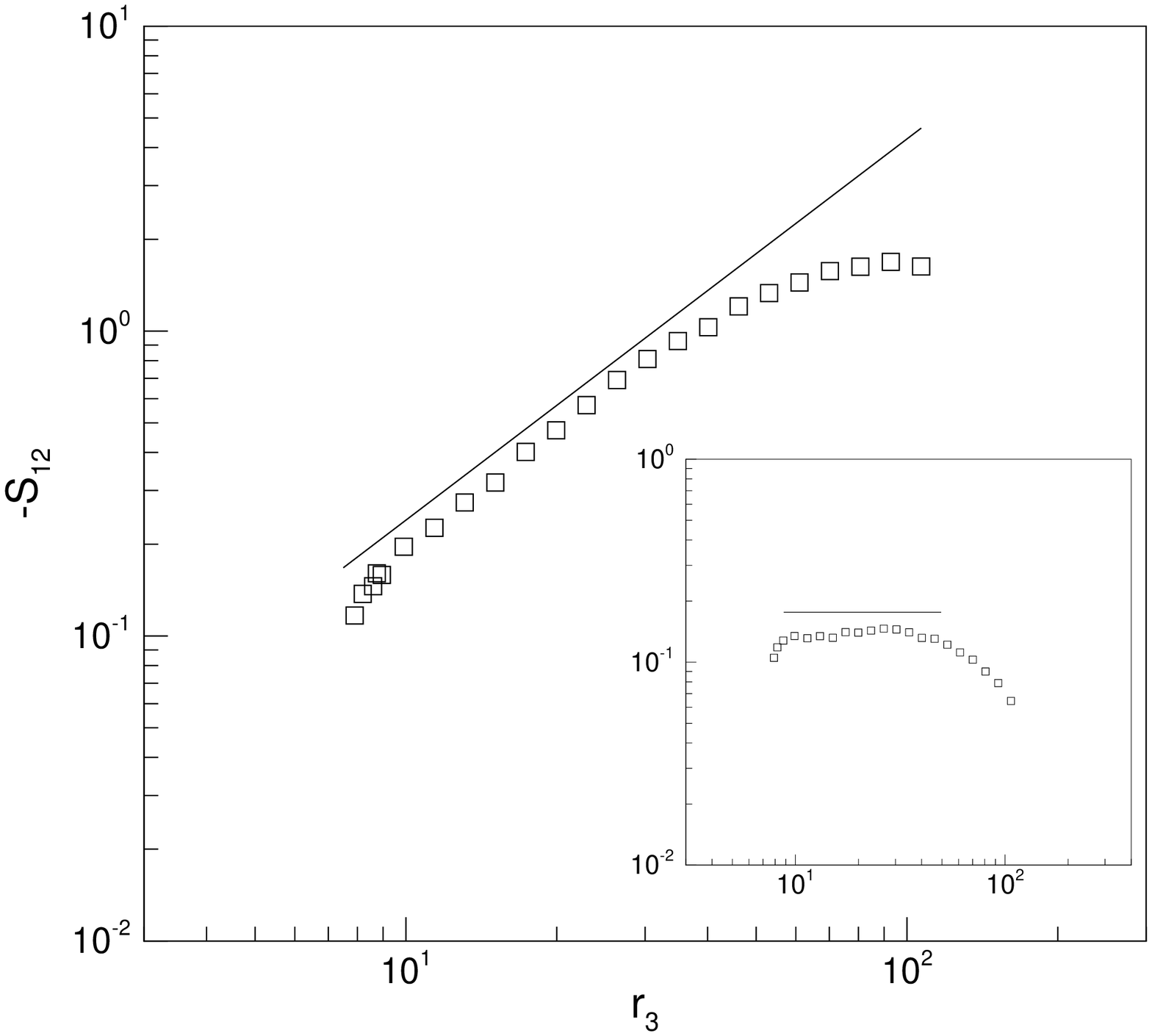,width=8.0cm}}}
{\centerline {
\epsfig{figure=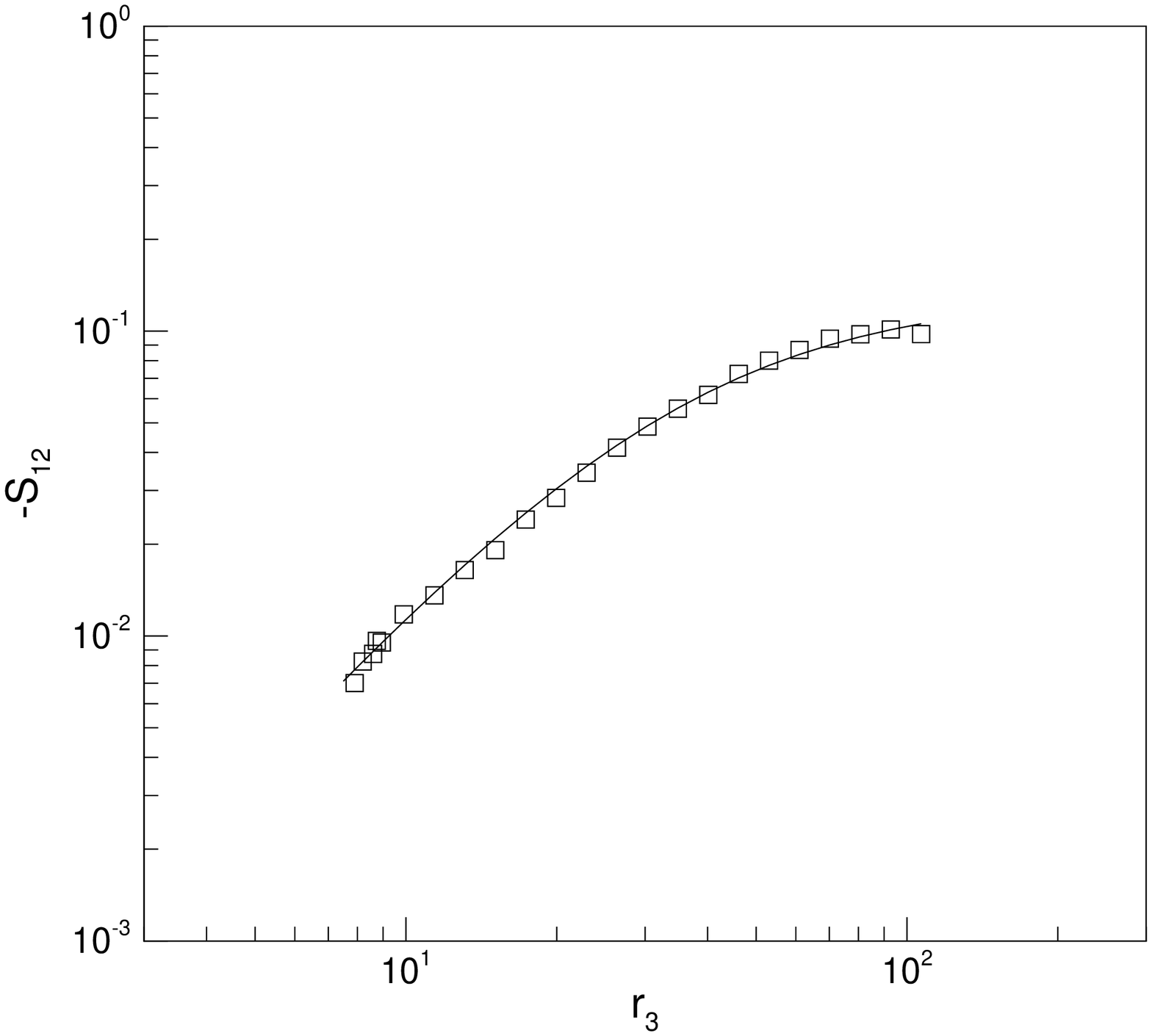,width=8.0cm}}}
   \caption{Homogeneous shear flow. Top: Log-log plot of the mixed structure
            function of second order,
            $-S_{12}$ plotted as a function of the transverse separation
            $r_3$. All quantities are normalized with respect to the inner
            scaling defined by $u_\tau = \sqrt{\tau/\rho}$ and $\nu/u_\tau$.
            The solid line corresponds to the slope $1.2$. The minimal length scale resolved is of the order of the dissipative scale.
            In the inset the same data are shown in compensated form,
            $-S_{12}/{r_3}^{1.2}$.
            Bottom:
            The same data are fitted by means of the expression given in
            equation~(\ref{eq:drhuva}), with $A_2 =0.018$, $B_2 =0.062$ and
            $D_2=3.27$,   solid line.
            \label{HSS12}
}
\end{figure}

In  figure~\ref{HSS12},  the second order mixed structure 
function $S_{12}(r_3)$ is plotted as a function of the transverse separation. 
In addition to the $r^2$ behavior at smaller scales and to the large scale 
saturation, a power-law at intermediate scales emerges distinctly, 
allowing us to measure the scaling exponent $\xi_2(2)$  with good 
accuracy. The estimate $\xi_2(2)=  1.2 \, \pm 0.07$ is indicated by the 
solid line, while the horizontal plateau in the inset, displaying the structure 
function in compensated form, shows the extension of the scaling 
region.

\begin{figure}[h]
{\centerline {
\epsfig{figure=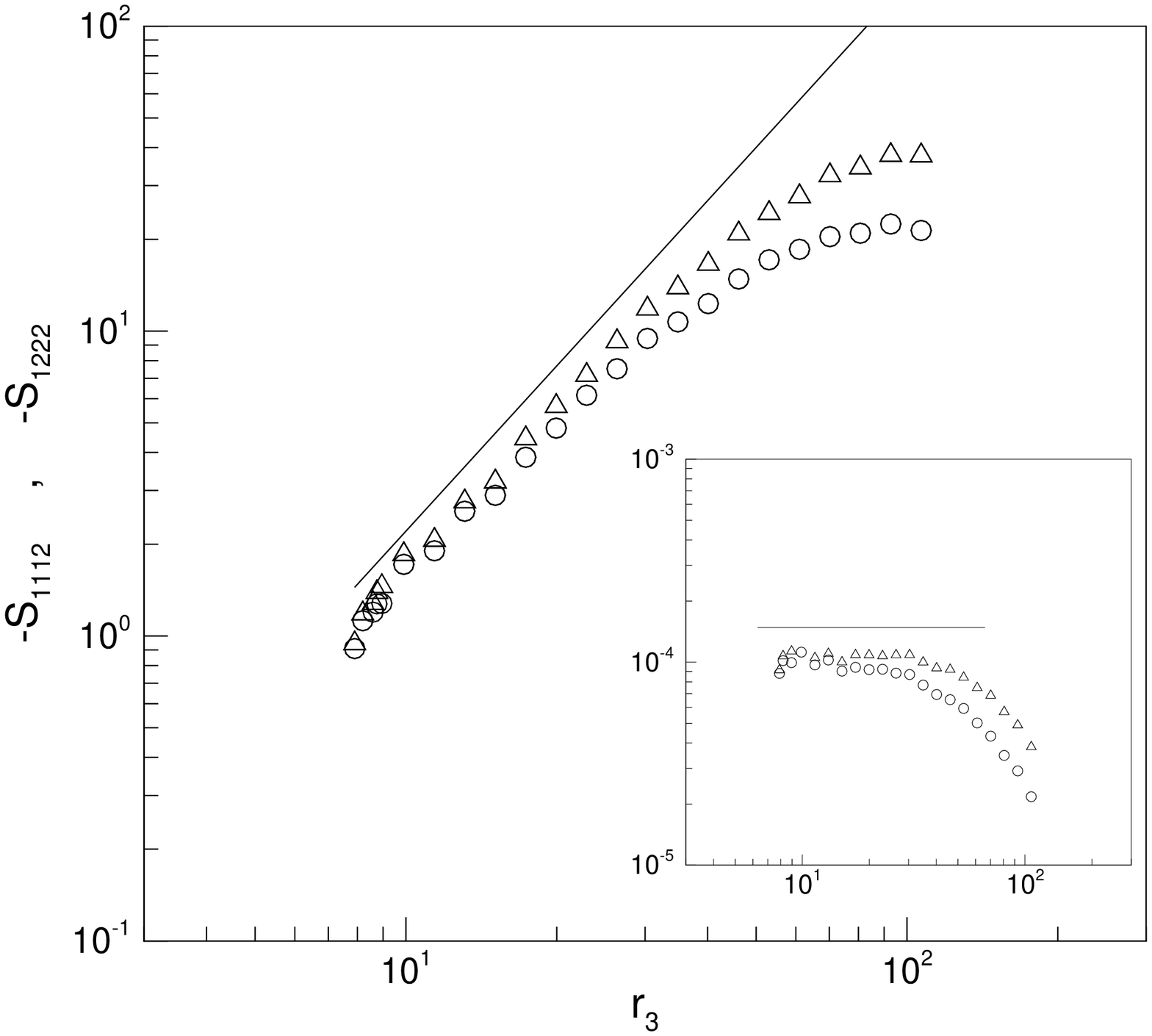,width=8.0cm}}}
{\centerline {
\epsfig{figure=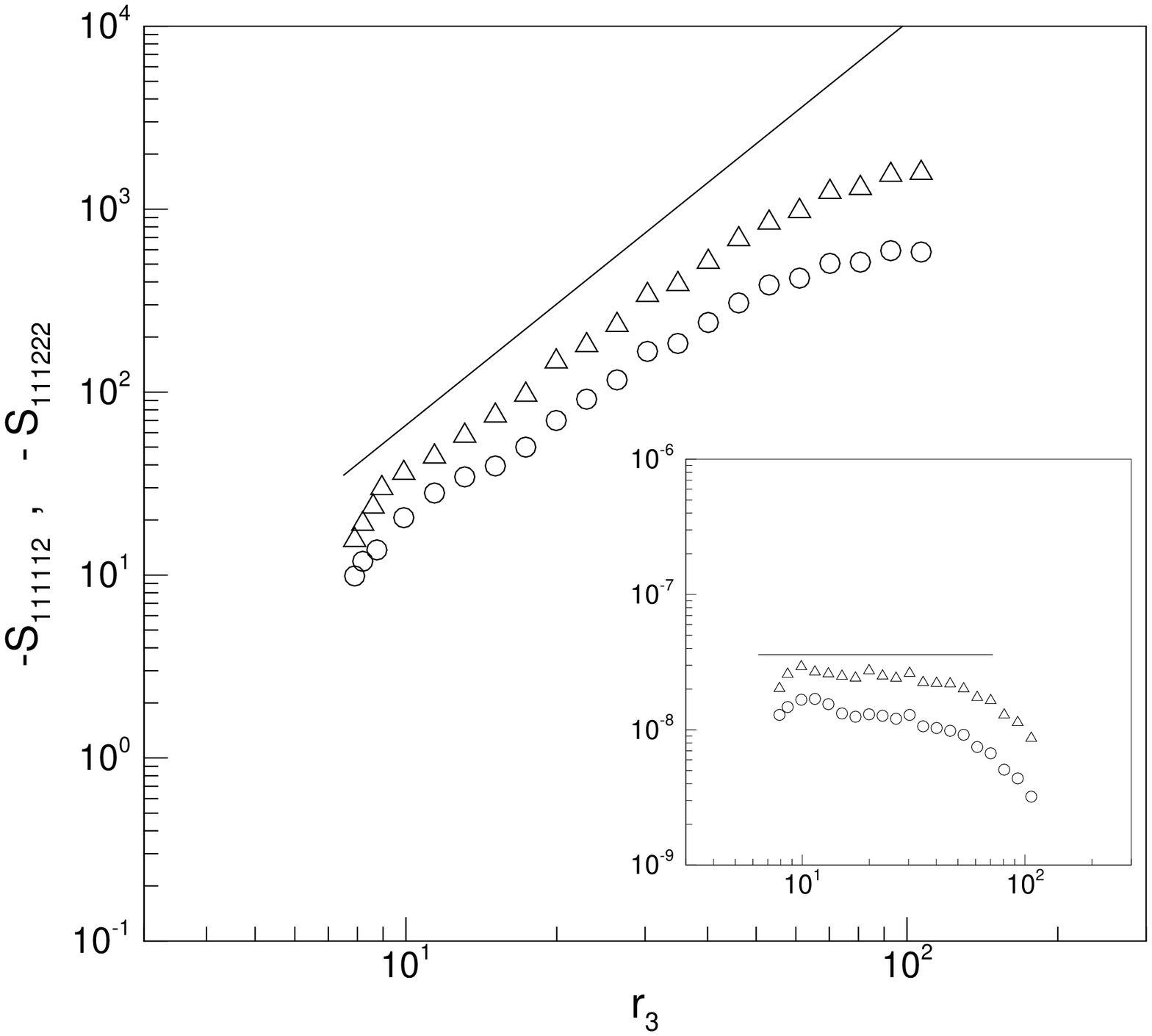,width=8.0cm}}}
   \caption{Homogeneous shear flow.
            Top panel:
            The purely anisotropic mixed structure functions of order
            $4$, $S_{1112}$ (triangles) and $S_{1222}$  (circles).
            The solid line indicates the slope of $1.7$.
            In the inset the same data are compensated with $r^{1.7}$, same symbols.
            Bottom plot:
            Two purely anisotropic mixed structure functions of order $6$,
            $S_{111112}$ (triangles) and  $S_{111222}$  (circles).
            The solid line indicates the slope of $2.05$.
            In the inset the same data are compensated with $r^{2.05}$, same symbols.
            \label{HS4}
}
\end{figure}

The same data  have also been  fitted by means of the expression proposed 
in \cite{kur00} to model the behavior of a generic 
structure function in the entire range of scales (bottom panel, solid line). 
In our case, the interpolation function for $S_{12}(r_3)$ is given by
\bea
\label{eq:drhuva}
S_{12}(r_3) \, = \, 
\frac{ A_2 \,  r_3^2 \, 
\left[\,1\,+\,D_2\,\left(\,r_3/\ell_{\rm T}\,\right)\,
\right]^{-  \xi_2(2)} 
}
{\left[\,1\,+\,B_2\,\left(\,r_3/\eta\,\right)^2 \, 
\right]^{1-\left[{ \xi_2(2)}\right]/2}  \, } \, . 
\eea
\noi 
and describes the superposition of a scaling behavior with coefficient 
$\xi_2(2)$ at intermediate scales, a  large-scale
saturation and a  dissipative closure at small scales.   
Here, the exponent $ \xi_2(2)$ is fixed by the direct fit estimated from the 
compensated plot in the top panel of figure~\ref{HSS12}, the 
transverse integral scale $\ell_{\rm T}$ is evaluated according to its 
definition given in table~\ref{tab:HS-table}, while $A_2$ and $B_2$ are the 
only two fitting constants. 
The ability of equation~(\ref{eq:drhuva}) to correctly reproduce the 
experimental data can be appreciated by looking at  the excellent agreement 
between the solid curve and the open symbols in the bottom panel of 
figure~\ref{HSS12}.  Such an extra fitting procedure, not strictly necessary 
here, turns out to be useful later in the context of the TBL flow.

Results concerning higher-order statistics of anisotropic fluctuations 
are reported in figure~\ref{HS4}.  
In the  top panel, the two  transverse observables of  order four, namely
$S_{1112}(r_3)$ and  $S_{1222}(r_3)$ are shown, both in their standard and 
compensated forms. 
A best fit  yields for the  exponent $\xi_2(4)$ a value of  $1.7 \pm 0.1$, 
indicated by the solid line. 
The  associated error accounts  for  both  the deviation from a pure  scaling law  and for the slightly  different behavior of 
$S_{1112}(r_3)$ with respect to  $S_{1222}(r_3)$.  
The corresponding compensated 
structure functions of order four are shown in  the inset of the top  panel.  
Mixed structure functions of order six are displayed in 
the bottom panel of figure~\ref{HS4}. In particular, with the configuration of 
figure~\ref{2sondeX}, three transverse observables can be measured,  
namely $S_{111112}(r_3)$, $S_{111222}(r_3)$  and  $S_{122222}(r_3)$.  
Only the first two, with the best statistical properties, are shown in the figure. 
Here again, minimal differences in the scaling behavior of these quantities are observed. Typically, 
lower values of the exponents are achieved for structure functions with largest weight on the vertical velocity
 component, i.e. for the sixth order $S_{122222}$.
The exponents increase with increasing weight of $u$, i.e. for the
sixth order moving from $S_{122222}$ through  $S_{111222}$ to  $S_{111112}$.
However, we find that a unique value of  $\xi_2(6) = 2.05 \, \pm 0.15 $ is able to fit satisfactorily
the set of statistics of order  six.  

The scaling exponents extracted by the present procedure
are  consistent with the values given in \cite{she02a} and in \cite{kur00}, 
namely $\xi_2(2)= 1.05 \div 1.22$, $\xi_2(4) = 1.42 \div 1.56$ and  
$ \xi_2(6) = 1.71 \div 2.14$.

%
\subsection{The turbulent boundary layer}
%
\label{subsec:BL} 

To address the effect of the shear intensity and of different boundary 
conditions, we consider the more complex environment represented by  
the near-wall region of a fully developed turbulent boundary layer. 
In this flow configuration two basic difficulties emerge.
The first one is associated with the thinness of the region where significant
changes of the mean gradient occur which poses severe restrictions on the
probe dimensions.
The second problem is related to the relatively large fluctuation level
in the lower part of the log-region, which may cause troubles with Taylor 
hypothesis.  
Concerning the first issue, a boundary layer as thick as possible was realized 
in a relatively large, yet well-controlled, experimental facility, as 
described in section~II. 
\begin{figure}[b!]
{\centerline {
\epsfig{figure=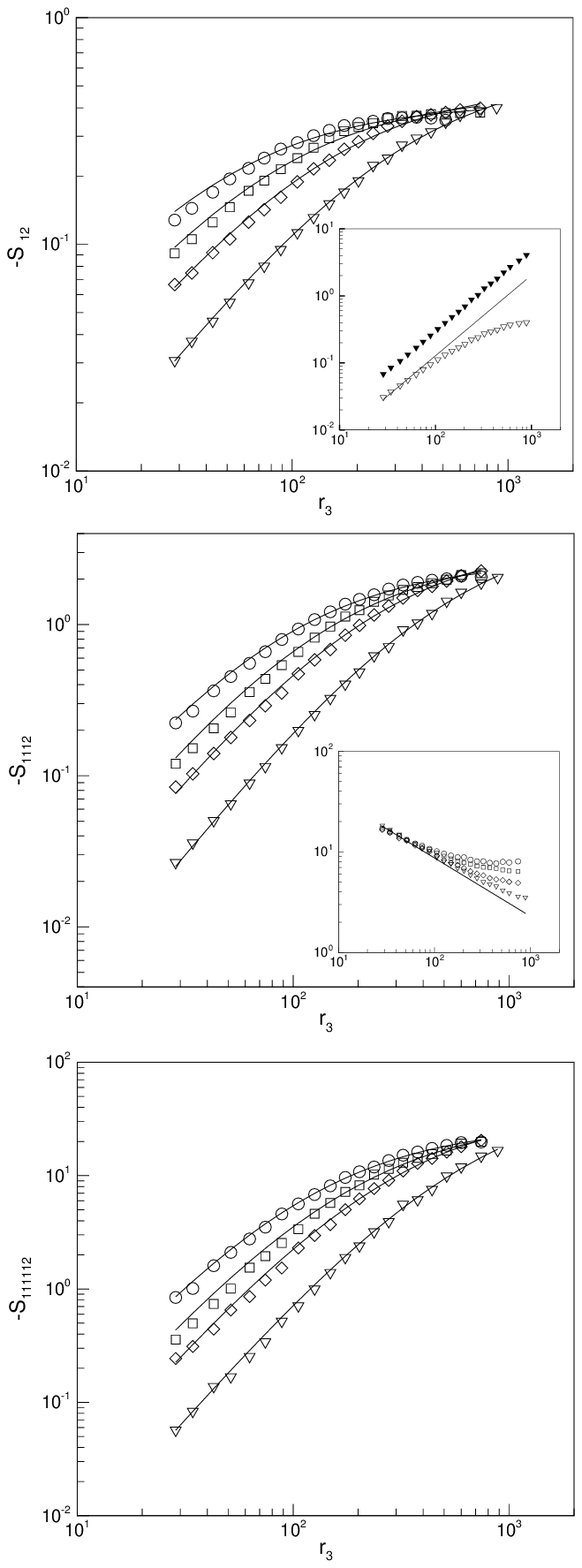,width=8.cm}}}
\end{figure}
\begin{figure}[t!]
 \caption{ Turbulent boundary layer:
            log-log plot of the
            mixed structure functions vs transverse
            separation $r_3$ at different wall distances
            $y^+ = 350$ (triangles), $240$ (diamonds), $140$ (squares),
            $90$ (circles) and for different orders $n = 2, \, 4,\, 6$.
            In the top panel we show $-S_{12}$  (second order) where
            the corresponding solid lines represent the
            fit~(\ref{eq:drhuvab})
            keeping fixed
            $ \xi_2(2) = 1.2$  at varying $A_2,  \,  D_2$.
            The inset shows $-S_{12}$ in raw form (open triangles)  and after
            compensation with saturation and viscous contributions
            (filled triangles) at $y^+ = 350$.
            The solid line indicates the slope $1.2$.
            In the middle panel we show $-S_{1112}$  (fourth order),
            with $ \xi_2(4) = 1.7$  at varying $A_4, \,  D_4$.
            In the inset we show the anisotropic flatness
            $-S_{1112}/S_{12}^2$ at corresponding
            locations.  The solid line indicates the slope $-.65$.
            In the bottom panel  we show  $-S_{111112}$ (sixth order)
            with $ \xi_2(6) = 2.05$ at varying $A_6,  \,  D_6$.
            \label{TBLS12}
                           }
\end{figure}

Moreover, to ensure a sufficient probe resolution, the measurements were 
limited to the log-region,  sufficiently far  from the boundary. 
The second point was instead entirely by-passed by addressing anisotropic
observables depending only on the transverse separation $r_3$, as already 
explained for the HS configuration.
Also for the boundary layer, the data we discuss are presented in 
dimensionless form, using the friction velocity

\bea
\label{u_tau_BL}
u_\tau \, = \, \sqrt{ \tau_w /\rho} \, 
\eea

\noi as characteristic velocity scale, where $\tau_w$ is the average shear 
stress at the wall (section~\ref{apparatus}). 
This corresponds directly to the normalization used in 
subsection~\ref{subsec:HS}, since in the near-wall region the total shear 
stress $\tau$ is constant in the wall-normal direction.
The main issue is connected to the assessment of  the anisotropic properties 
at changing distance from the wall.

In figure~\ref{TBLS12} we summarize the results for the fully anisotropic
transverse structure functions already introduced in the previous section.  
The three panels show, from top to bottom, the observables of order 2, 4 and 6, 
respectively, while the different symbols correspond to  different distances 
from the wall, from $y^+ = 350$ down to $y^+=90$. 
As one can see, the scaling properties are not 	as clear as in the HS case.
Independently of the order of the structure function, farther from the wall
a distinct scaling range emerges. As the wall is approached, the scaling
behavior is less evident and a tendency towards saturation at large scales
is observed. 
Here, in order to extract quantitative results one needs to consider also
large scale effects.
In particular, we generalized the expression~(\ref{eq:drhuva}) to all orders,
for separations much larger than the Kolmogorov scale:
\bea
\label{eq:drhuvab}
S_{\alpha_1 \ldots \alpha_n}(r_3) \, = \, 
{ A_n(y^+) \,  r_3^{\xi_2(n)} \, 
\left[\,1\,+\,D_n(y^+)\,\frac{\ds r_3}{\ds \ell_{\rm T}}\,
\right]^{-  \xi_2(n)} 
}
 \, . 
\eea
By comparing the fit with the raw data, it is quite clear that the poor 
scaling closer to the wall is substantially explained in terms of 
saturation occurring earlier and earlier as the wall is approached.
The inset in the top panel describes the fitting procedure and highlights the 
scaling law by removing the effect of the large scale saturation.
The inset of the middle panel shows the  anisotropic flatness, $-S_{1122}/S_{12}^2$, to highlight the high degree of intermittency showed by anisotropic fluctuations, independently on the distance from the wall. 
It is important to stress here that the good agreement with the data for all
distances is obtained by keeping fixed  the scaling exponents $\xi_2(n)$ to the value
obtained in the HS case for all orders $n$ --only prefactors change at changing
the wall distance.

%
%
%
%

\section{Concluding Remarks}
\label{concluding}
In this paper we have performed a systematic analysis of small-scales
anisotropic turbulent properties in two different experimental set-up,
a homogeneous shear and a turbulent boundary layer. We have used two
cross-wires probes in order to extract the leading anisotropic
fluctuations of two-point correlation functions in the homogeneous
directions without the needing of Taylor hypothesis. We have analyzed
structure functions up to order $n=6$ finding a good agreement of the 
anisotropic exponents between the two experimental set-up and at
changing the distance from the wall in the turbulent boundary layer.
We could compare the anisotropic properties by changing the normalized
shear intensity, $S^*$, by a factor 2 and more.\\ The cleanest data
are obtained for the homogeneous shear flow  where a fit
of the power law behavior allowed for a direct measurement of the
anisotropic properties. In the turbulent boundary layer, we had to
take into account also large scale saturation effects, especially
close to the wall, in order to obtain a global fit of the structure
functions behavior for all value of $y^+$. Our results support the
``exponent-only'' scenario of universality also in the anisotropic
sector. In other word, we have been able to fit all experimental data
by keeping fixed the scaling properties and adjusting only the
prefactors.\\   The main drawback of all
actual experimental set-up consists in the impossibility to exactly
disentangling different anisotropic sectors among themselves. This
implies that sub-leading contributions coming from sector with higher
$j$'s could spoil the quantitative measurements. In particular, the
systematic differences observed here and in other studies \cite{she02a,kur00}
between the scaling of anisotropic correlation functions of the same
order but with different tensorial components as $S^{(6)}_{111222}(r)$
and $S^{(6)}_{111112}(r)$ may well be due to the effects of sub-leading
contributions coming from the $j \ge 4$ sectors in the ${\rm SO}(3)$ expansion.
%
%
%
%
\begin{acknowledgments}
We acknowledge useful discussions with R. Piva and S. Kurien.
This research was supported by MIUR and the EU under the Grants
No. HPRN-CT 2000-00162 ``Non Ideal Turbulence''.
\end{acknowledgments}

%
%

%
%


\begin{thebibliography}{99} 
\bibitem{fri95}
U. Frisch, 
{\em Turbulence: The legacy of A.N. Kolmogorov} 
(Cambridge University Press, Cambridge, 1995).  
\bibitem{gar98}
S. Garg and Z. Warhaft, 
``On the small scale structure of simple shear flow",
 Phys.~Fluids {\bf 10} (3),  662-673,  (1998).
\bibitem{she00}
X. Shen and Z. Warhaft, 
``The anisotropy of the small scale structure in high Reynolds
 number ($R_\lambda \sim 1000$) turbulent shear flow",
Phys.~Fluids {\bf 12} (11),  2976-2989,  (2000).
\bibitem{she02a}
X. Shen and Z. Warhaft, 
``On the higher order mixed structure functions in laboratory
 shear flow", 
Phys.~Fluids {\bf 14} (7),  2432-2438,  (2002).
\bibitem{pum96}
A. Pumir, 
``Turbulence in homogeneous shear flows", 
Phys.~Fluids 
{\bf 8} (11), 3112-3127,  (1996).
\bibitem{bif01a}
L. Biferale and F. Toschi, 
``Anisotropic homogeneous turbulence: Hierarchy and
 intermittency of scaling exponents in the anisotropic sectors", 
Phys. Rev. Lett. {\bf 86} (21),  4831-4834,  (2001).
\bibitem{bif01}
L. Biferale and M. Vergassola, `
`Isotropy vs anisotropy in small-scale turbulence", 
Phys.~Fluids {\bf 13} (8),  2139-2141,  (2001).
\bibitem{ara99b}
I. Arad, V. L'vov, and I. Procaccia, 
``Correlation functions in isotropic and
 anisotropic turbulence: The role of the symmetry group", 
Phys. Rev. E {\bf 59} (6),  6753-6765,  (1999).  
\bibitem{ara99}
I. Arad, L. Biferale, I. Mazzitelli, and I. Procaccia,
``Disentangling scaling properties in anisotropic and inhomogeneous turbulence",
  Phys. Rev. Lett. {\bf 82} (25),  5040-5043,  (1999).
\bibitem{ara98}
I. Arad, B. Dhruva, S. Kurien, V.~S. L'vov, I. Procaccia, and K.~R.
  Sreenivasan, 
``Extraction of anisotropic contributions in turbulent flows", 
Phys.~ Rev. Lett. {\bf 81} (24),  5330-5333,  (1998).  
\bibitem{kur00}
S. Kurien and K.~R. Sreenivasan, 
``Anisotropic scaling contributions to high-order
 structure functions in high-Reynolds-number turbulence", 
Phys.~ Rev. E {\bf 62} (2),  2206-2212,  (2000).  
\bibitem{kur00a}
S. Kurien, V. L'vov, I. Procaccia, and K. Sreenivasan,
 ``Scaling structure of the velocity statistics in atmospheric boundary layers",
Phys. Rev. E {\bf 61} (1), 407-421, (2000).
\bibitem{bif02} 
 L. Biferale, I. Daumont, A. Lanotte  and F. Toschi  
``Anomalous and Dimensional scaling in anisotropic turbulence''
{ Phys. Rev. E} {\bf 66} (5), 056306, (2002).
\bibitem{dec}
 L. Biferale, G. Boffetta, A. Celani, A. Lanotte, F. Toschi and M.
 Vergassola; ``The decay of homogeneous anisotropic turbulence'', 
 {\it Phys. Fluids} {\bf 15} (8)  2105,  (2003).
\bibitem{arneodo} 
A. Arneodo {\em {et al.}}, 
``Structure functions in turbulence, in various flow configurations, 
at Reynolds number between 30 and 5000, using extended self-similarity", 
Europhys. Letters {\bf {34}} (6), 411-416, (1996).  
\bibitem{wandervater} 
A. Staicu, B. Vorselaars, and W.  van de Water,
``Turbulence anisotropy and the SO(3) description"
 Phys. Rev. E {\bf {68}} (4) (2003). 
\bibitem{bif03a}
L. Biferale, E. Calzavarini, F. Toschi, and R. Tripiccione,
``Universality of anisotropic fluctuations from numerical simulations
 of turbulent flows", 
Europhys. Lett. {\bf {64}} (4),461-467, (2003).
  (2003).
\bibitem{Mazzitelli}
L. Biferale, D. Lohse,  I.M. Mazzitelli and F. Toschi, 
``Probing structures in channel flow through SO(3) and SO(2) decomposition"
J. Fluid Mech 452, 39-59, (2002). 
\bibitem{Tavoularis} 
S. \@ Tavoularis,and S.\@  Corrsin,   
{``Experiments in nearly homogeneous turbulent shear flow with
 a uniform mean temperature gradient. Part 1"}, 
J. Fluid Mech. {\bf 104}, 311-348, (1981);
``Experiments in nearly homogeneous turbulent shear flow with
 a uniform mean temperature gradient. Part 2. The fine structure", 
{\em {ibid.}} {\bf 104}, 349, (1981).
\bibitem{Meneveau}
S.\@ Cerutti and Ch.\@  Meneveau,  
{``Statistics of filtered velocity in grid and wake turbulence"}, 
Phys. Fluids {\bf 12} (5), 1143-1162,  (2000).
\bibitem{lumley}
   J\@. L\@. Lumley,
 ``Interpretation of time spectra measured in high-intensity shear flows",
 Phys. Fluids, {\bf 8} (6), 1056-1062, (1965).

\end{thebibliography}
\end{document}